\documentclass[twocolumn,english]{revtex4-1}
\usepackage[T1]{fontenc}
\usepackage[latin9]{inputenc}
\usepackage{amstext}
\usepackage{graphicx}
\usepackage{babel}
\begin{document}

\title{Hard X-ray spectroscopy of the itinerant magnets $R$Fe$_{4}$Sb$_{12}$
($R=$Na, K, Ca, Sr, Ba)}

\author{Bassim Mounssef Jr.$^{1}$, Marli R. Cantarino$^{2}$, Eduardo M.
Bittar$^{3}$, Tarsis M. Germano$^{2}$, Andreas Leithe-Jasper$^{4}$,
 Fernando A. Garcia$^{2}$}

\affiliation{$^{1}$IQUSP, Univ. de São Paulo, 05508-090, São Paulo-SP, Brazil}

\affiliation{$^{2}$IFUSP, Univ. de São Paulo, 05508-090, São Paulo-SP, Brazil}

\affiliation{$^{3}$Centro Brasileiro de Pesquisas Físicas, Rio de Janeiro, RJ
22290-180, Brazil}

\affiliation{$^{4}$Max Planck Institute for Chemical Physics of Solids, D-01187
Dresden, Germany.}
\begin{abstract}
Ordered states in itinerant magnets may be related to magnetic moments
displaying some weak local moment characteristics, as in intermetallic
compounds hosting transition metal coordination complexes. In this
paper, we report on the Fe $K$-edge X-ray absorption spectroscopy
(XAS) of the itinerant magnets $R$Fe$_{4}$Sb$_{12}$ ($R=$Na, K,
Ca, Sr, Ba), aiming at exploring the electronic and structural properties
of the octahedral building block formed by Fe and the Sb ligands.
We find evidence for strong hybridization between the Fe $3d$ and
Sb $5p$ states at the Fermi level, giving experimental support to
previous electronic structure calculations of the $R$Fe$_{4}$Sb$_{12}$
skutterudites. The electronic states derived from Fe 3$d$ Sb $5p$
mixing are shown to be either more occupied and/or less localized
in the cases of the magnetically ordered systems, for which $R=$
Na or K, connecting the local Fe electronic structure to the itinerant
magnetic properties. Moreover, the analysis of the extended region
of the XAS spectra (EXAFS) suggests that bond disorder may be a more
relevant parameter to explain the suppression of the ferromagnetic
ordered state in CaFe$_{4}$Sb$_{12}$ than the decrease of the density
of states.
\end{abstract}
\maketitle

\section{Introduction}

Magnetic ordered states in itinerant magnets are understood to be
connected to a magnetic instability due to a high density of states
at the Fermi surface, thus being a property of itinerant electronic
states. This is in contrast to the magnetic properties of insulators
which stems from localized electronic states \cite{white_quantum_2007,kubler_theory_2009}.

Although this formal division of magnetism from either itinerant or
localized moments offers an important rationalization of the problem,
it is striking that there are only few solids hosting magnetic ordered
states for which the magnetic properties bears no relation whatsoever
to atomic electronic properties. To our knowledge, examples include
ZnZr$_{2}$ \cite{matthias_ferromagnetism_1958}, ScIn$_{3}$ \cite{matthias_ferromagnetism_1961}
and the recently discovered AuTi \cite{svanidze_itinerant_2015}.
In general, magnetic moments in itinerant magnets may display weak
local moment characteristics \cite{kubler_theory_2009}.

The iron based superconductors \cite{kamihara_iron-based_2008} illustrates
an example of itinerant magnets for which the specific Fe orbital
configuration, valence and coordination structure are important parameters.
In this broad research field, the presence of a structural building
block, featuring Fe atoms in tetrahedral coordination, is a common
thread. The experimental investigation by X-ray absorption spectroscopy
(XAS) of the electronic and structural properties of this coordination
structure brought into light several questions in the field, including
the role of transition metal substitution \cite{bittar_co-substitution_2011,merz_electronic_2012,baledent_electronic_2015},
bond disorder \cite{granado_pressure_2011,cheng_overview_2014,cheng_charge_2015,hacisalihoglu_study_2016}
and the relevance of Mott physics \cite{lafuerza_evidence_2017}. 

As opposed to the case of local moment magnets \cite{white_quantum_2007},
it is not clear at which situation the electronic and structural properties
of a coordination structure will be relevant to the magnetic properties
of itinerant magnets. In fact, it is expected that in many situations
the conduction electrons will completely screen the effect of the
ligands \cite{kubler_theory_2009}. The experimental investigation
of itinerant magnets hosting coordination structures may challenge
this understanding, opening a new perspective to the study of itinerant
magnets. 

\begin{figure}
\begin{centering}
\includegraphics[scale=0.28]{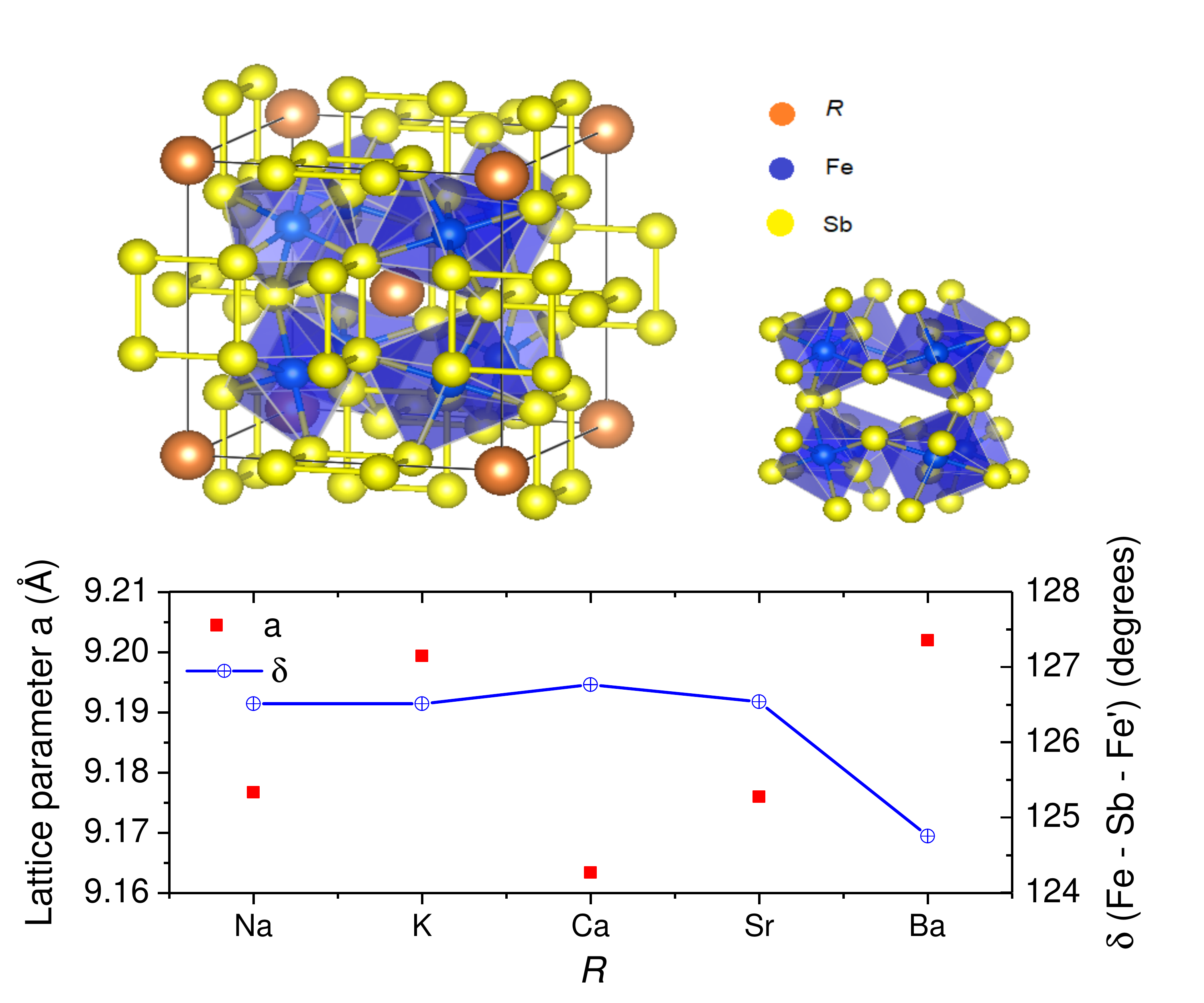}
\par\end{centering}
\caption{(Color online) Left: Ternary filled skutterudite structure (space
group $Im\bar{3}$) displaying the rectangular Sb-Sb rings and  the
FeSb$_{3}$ distorted octahedra \cite{momma_vesta:_2008}. Right:
detail of the Fe-Sb octahedral coordination \label{fig:structure_properties}.
Below: the lattice parameters $a$ (red squares) and the Fe - Sb -
Fe' bond angles $\delta$ (blue line) along the series. Is is noteworthy
that in spite of the large variation of the cationic sizes, $a$ changes
only slightly along the series, evidencing that the Fe$_{4}$Sb$_{12}$
framework is particularly stiff.}

\end{figure}

In this paper, we present Fe $K$-edge $X$-ray absorption spectroscopy
(XAS) of the $R$Fe$_{4}$Sb$_{12}$ ($R=$Na, K, Ca, Sr, Ba) filled
skutterudites. In the filled skutterudite structure \cite{jeitschko_lafe4p12_1977},
the metal ($M$) is coordinated by $6$ ($X$) ligands (mainly $X=$P,
As, Sb, although $X=$Ge is also possible), arranged in an octahedral
geometry (see Fig. \ref{fig:structure_properties}). Our Fe $K-$edge
XAS experiments allowed a detailed study of the structural and electronic
properties of the FeSb$_{6}$ building blocks along the $R=$Na, K,
Ca, Sr, Ba series. 

This series of $R$Fe$_{4}$Sb$_{12}$ filled skutterudites presents
an interesting case study of skutterudite compounds for which electronic
and magnetic properties are dominated by the transition metal magnetism
\cite{schnelle_itinerant_2005,matsuoka_nearly_2005,berardan_rare_2005,sichelschmidt_optical_2006,takabatake_roles_2006,yamaoka_strong_2011}.
A high density of states at the Fermi level ($E_{F}$) is inferred
based upon the measured electronic contribution to the heat capacity
$\gamma$, presenting values of the order of $\approx100$ mJmol$^{-1}$K$^{-2}$.
Electronic structure calculations attribute the presence of such heavy
quasiparticles at $E_{F}$ to strongly hybridized Fe - Sb states at
$E_{F}$. The alkaline metal filled skutterudites ($R=$ Na, K) present
a ferromagnetic phase transition at the critical temperature ($T_{C}$)
about $\approx80$ K, whereas the alkaline earth filled ($R=$ Ca,
Sr, Ba) remain paramagnetic for temperatures down to $T=2$ K \cite{leithe-jasper_ferromagnetic_2003,leithe-jasper_weak_2004,schnelle_magnetic_2008}. 

A qualitative understanding about the evolution of the magnetic properties
may be formulated in terms of the electron filling along the series.
In reference to the fully compensated semiconductor (CoSb$_{3}$)$_{4}$,
the (FeSb$_{3}$)$_{4}$ framework is short of $4$ electrons. The
stabilization of the structure is provided by electron transfer from
the filler cation $R$ \cite{jeitschko_lafe4p12_1977,noauthor_filled_2010}.
However, the monovalent alkaline metal or divalent alkaline earth
fillers can only partially compensate the host structure. Charge carriers
in $R$Fe$_{4}$Sb$_{12}$ are, therefore, holes and, due to the extra
electron, the alkaline earth filled systems present a smaller density
of states at the Fermi level ($E_{F}$). In turn, the Stoner factor
should be less for the systems for which $R=$ Ca, Sr, or Ba, suppressing
the magnetic instability that gives rise to ferromagnetic order in
the $R=$ Na or K systems \cite{schnelle_magnetic_2008}. 

Electronic structure calculations do support this qualitative scenario,
favoring an itinerant description of the magnetic properties of these
skutterudites \cite{schnelle_magnetic_2008}. On the other hand, neutron
scattering experiments of NaFe$_{4}$Sb$_{12}$ could demonstrate
the presence of local moments at the Fe sites \cite{leithe-jasper_neutron_2014}
and further investigation of the electronic structure of NaFe$_{4}$$X$$_{12}$
($X=$P, As, Sb) \cite{xing_magnetism_2015} revealed that the specific
nature of the ligand orbitals ($3p$ for P, $4p$ for As or $5p$
for Sb) implies a relevant effect in the size of the magnetic moment
and the nature of the magnetic instability. These two results suggest
that local properties of the FeSb$_{6}$ coordination structure could
relate to the evolution of the itinerant magnetism of the system.

\section{Methods}

High quality polycrystalline samples of $R$Fe$_{4}$Sb$_{12}$ ($R=$Na,
K, Ca, Sr, Ba) skutterudites were synthesized by the solid state method
\cite{leithe-jasper_ferromagnetic_2003}. Hard $X$-ray absorption
spectroscopy experiments were performed at the XAFS2 \cite{figueroa_upgrades_2016}
beamline of the Brazilian Synchrotron Light Source (CNPEM-LNLS). At
the XANES region, the spectra were measured in steps of $0.3$ eV
in both fluorescence and transmission mode for the light filler elements
($R=$ Na, K, Ca) whereas for the heavy fillers ($R=$ Sr or Ba) only
the fluorescence mode was considered. The fluorescence signal was
recorded by $11$ Ge detectors and averaged to improve the sinal-to-noise
ratio. An Fe foil, kept at room temperature, was measured in the transmission
mode as a reference throughout the experiments. A conventional He-flow
cryostat was employed to achieve temperatures down to $T=10$ K. Each
spectrum was measured $3$ times. 

Our interpretation of the results of the near edge region of the XAS
spectra (XANES) is supported by\emph{ Ab initio} calculations of multiple
scattering theory implemented by the FEFF$8.4$ code \cite{rehr_theoretical_2000}.
The\emph{ Ab initio} calculations were performed adopting clusters
of $226$ atoms and the Hedin-Lundqvist pseudopotential to account
for the exchange interaction. Self consistent calculations were performed
for a cluster radius of $6$ $\mathring{\text{A}}$. The calculations
of the spectrum include only contributions of dipolar nature. The
inclusion of quadrupolar contribution rendered a minute contribution
to the pre-edge region, which was disregarded. Cluster size effects
were investigated adopting NaFe$_{4}$Sb$_{12}$ as a reference and
no changes were observed for larger cluster sizes (up to $400$ atoms).
The effects of adopting larger clusters for the self consistency calculations
(up to $7$ $\mathring{\text{A}}$ ) were also tested but did not
result in more accurate results. For the extended region of the XAS
spectra (EXAFS), FEFF calculations were implemented along with IFEFFIT
using the Demeter platform \cite{ravel_athena_2005}. 

\section{Results and discussion}

\begin{figure}
\begin{centering}
\includegraphics[scale=0.33]{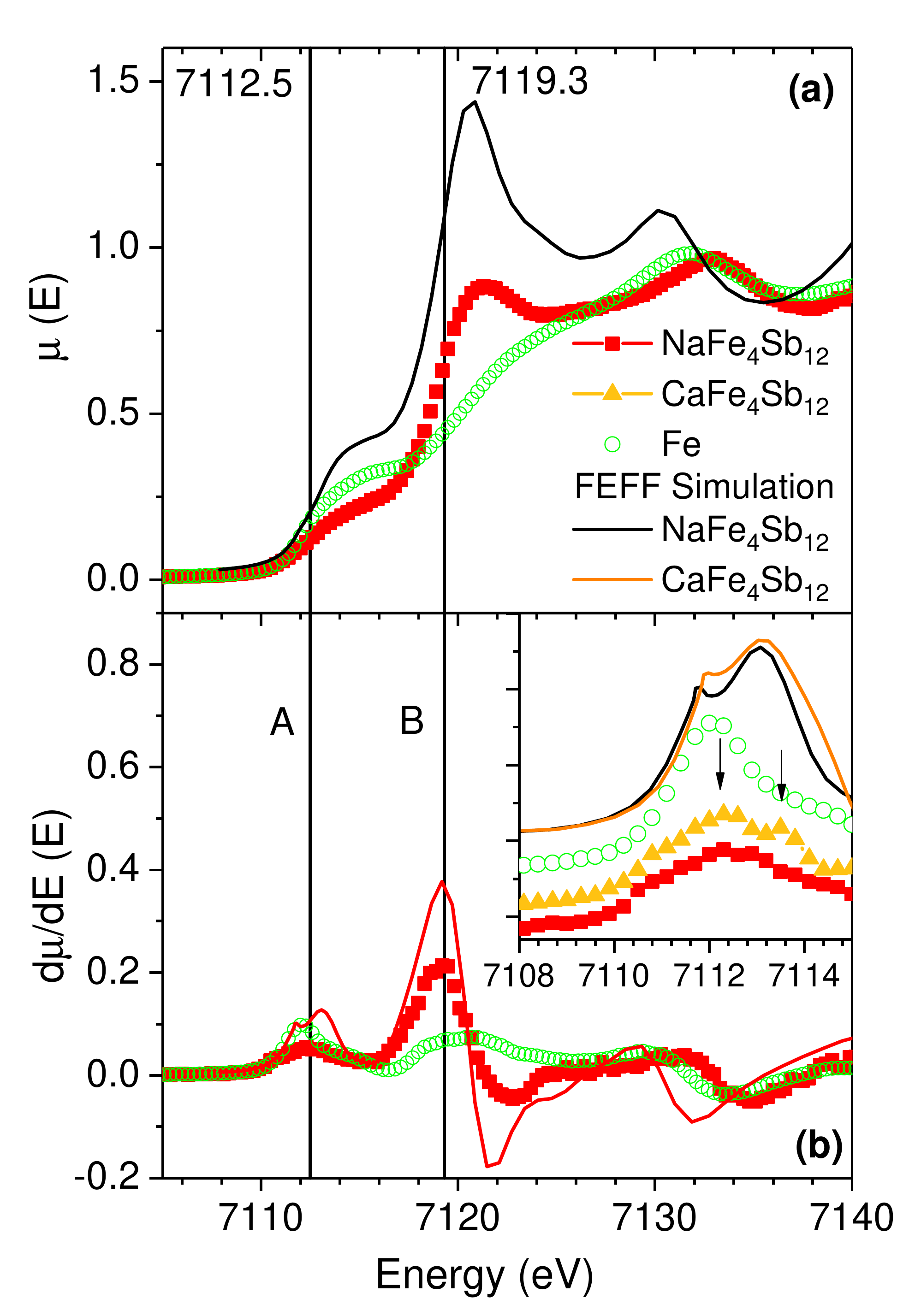}
\par\end{centering}
\caption{(Color online) $(a)$ Fe $K$-edge XANES spectra for NaFe$_{4}$Sb$_{12}$,
the reference Fe foil along with FEFF calculations for NaFe$_{4}$Sb$_{12}$,
and $(b)$ the derivative of the same spectra. Two prominent features
$A$ and $B$ are marked in reference to the maximum of the energy
derivatives of the spectrum. In the inset, the NaFe$_{4}$Sb$_{12}$
and CaFe$_{4}$Sb$_{12}$ spectra (off-set for better visualization)
are compared. It is shown that a second maximum (black arrows) of
the energy derivative can be observed for CaFe$_{4}$Sb$_{12}$ but
not for NaFe$_{4}$Sb$_{12}$. The overall shape of the spectrum is
reminiscent of the Fe $K$-edge spectra obtained for Fe$^{2+}$ oxides
in octahedral coordination complexes. \label{fig:basicXANES}}
\end{figure}

In Figs. \ref{fig:basicXANES} $(a)$-$(b)$ the Fe $K$-edge spectrum
of NaFe$_{4}$Sb$_{12}$ (XANES region), the respective FEFF simulation
and the spectrum of the Fe foil reference are presented (\ref{fig:basicXANES}
$(a)$) along with their energy derivatives (\ref{fig:basicXANES}
$(b)$). Capital letters $A$ and $B$ mark the position of the main
features of the NaFe$_{4}$Sb$_{12}$ spectrum, respectively the pre-edge
and edge transitions, determined by the inflection points of the spectrum.
To match the experimental transition edge position (feature $B$),
the FEFF calculation was shifted by $-4.3$ eV. Features $A$ and
$B$ are only qualitatively reproduced by the FEFF calculation. In
\ref{fig:basicXANES} $(a)$ it is shown that the simulation overestimates
the edge jump (see discussion below) and in \ref{fig:basicXANES}
$(b)$ it is suggested that the pre-edge transition splits in two
components which, however, are not observed for NaFe$_{4}$Sb$_{12}$. 

\begin{figure}
\begin{centering}
\includegraphics[scale=0.28]{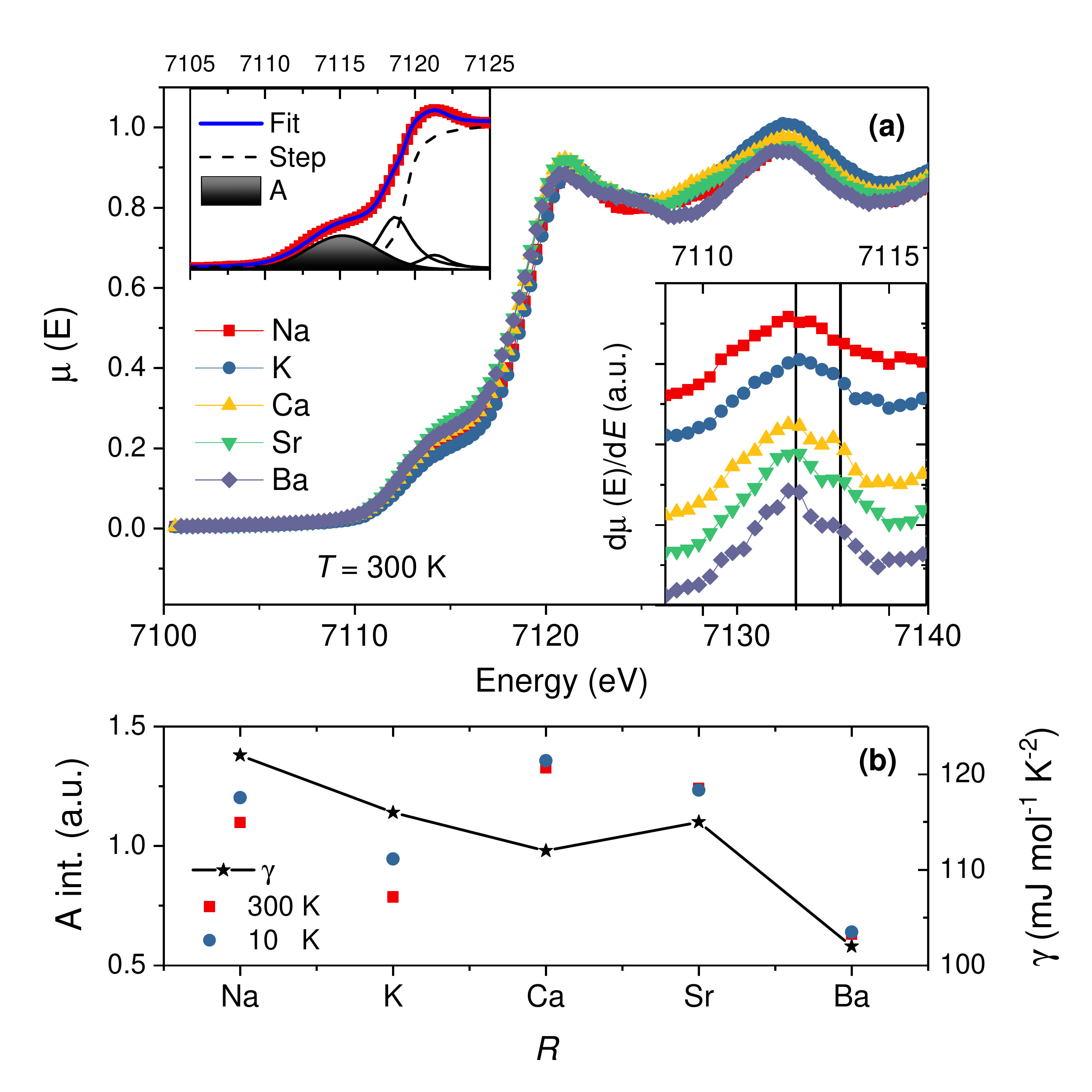}
\par\end{centering}
\caption{(Color online) $(a)$ XANES spectra for all $R$Fe$_{4}$Sb$_{12}$
($R=$Na, K, Ca, Sr, Ba) skutterudites. Upper inset: fitting of a
representative spectrum, wherein we highlight a Gaussian curve describing
the pre-edge. Lower inset: detail of the energy derivative of the
pre-edge transition for all samples. A second maximum can be observed
for the $R=$ Ca and likely for the $R=$ Sr, and Ba skutterudites.
$(b)$ Pre-edge transition intensity (peak area $A$, left axis) at
$T=300$ K (red squares) and $T=10$ K (blue squares) and the $\gamma$
coefficient (black line) of the electronic contribution to heat capacity
(right axis, from Ref.: \cite{schnelle_magnetic_2008}), both as a
function of $R$. \label{fig:XANESallDATA}}
 
\end{figure}

In the inset of Fig. \ref{fig:basicXANES} $(b)$, the energy region
of the pre-edge is presented in detail and the energy derivative of
the spectra for NaFe$_{4}$Sb$_{12}$ and CaFe$_{4}$Sb$_{12}$ are
compared with the respective FEFF simulations. For NaFe$_{4}$Sb$_{12}$,
the spectrum presents only one maximum and flattens, while for CaFe$_{4}$Sb$_{12}$
it presents a weak, however clear, second maximum (as indicated by
the black arrows). The FEFF calculated splitting of the pre-edge amounts
to $\Delta E^{\text{FEFF}}\approx1.2$ eV closely corresponding to
the observed value of $\Delta E^{R=\text{Ca}}\approx1.3(0.3)$ eV,
suggesting that the observed weak second maximum is an intrinsic property
of the system. The NaFe$_{4}$Sb$_{12}$ XANES spectrum compares well
with the $R=$ K spectrum while the pre-edge splitting observed at
the CaFe$_{4}$Sb$_{12}$ spectrum is also present for the $R=$ Sr
or Ba cases (see Fig. \ref{fig:XANESallDATA}$(a)$), although the
effect is weaker. 

Our FEFF simulations include only dipolar terms and thus we ascribe
the pre-edge to a dipolar transition. There are two mechanisms that
can account for a pre-edge intensity dominated by a dipolar transition,
and these are $i)$ a Fe $3d-4p$ mixing (allowing an onsite dipolar
Fe $1s\rightarrow3d$ transition) and $ii)$ a strong ligand $np$
(in this case Sb $5p$) mixing with the Fe $3d$ orbitals (allowing
an intersite Fe - Fe' $1s\rightarrow3d$ transition) \cite{debeer_george_metal_2005,yamamoto_assignment_2008,groot_1s_2009}.
If the Fe site is centrosymmetric, as in octahedral symmetry, the
Fe $3d-4p$ mixing is exactly suppressed. Therefore, the pre-edge
intensity should be taken as a piece of evidence for strong Fe $3d$
Sb $5p$ mixing.

Since the Fermi level ($E_{F}$) sits at the $A$ feature, experimentally
determined to be about $E_{F}\approx7112.5$ eV (for NaFe$_{4}$Sb$_{12}$),
our result offers compelling experimental evidence for a significant
contribution of Sb $5p$ orbitals at $E_{F}$. This contribution is
key for understanding the presence of heavy quasiparticles at $E_{F}$
in the $R$Fe$_{4}$Sb$_{12}$ skutterudites, that are directly connected
to their magnetic properties.

We caution, however, that in skutterudites the metal coordination
is not that of a regular octahedron and, in addition, may contain
a particularly large degree of disorder due to the unconventional
vibrational dynamics of these systems \cite{bridges_complex_2015,feldman_lattice-dynamical_2014,koza_vibrational_2011}.
Therefore, a contribution from Fe $3d-4p$ mixing to the pre-edge
cannot be entirely discarded.

\begin{figure}
\begin{centering}
\includegraphics[scale=0.28]{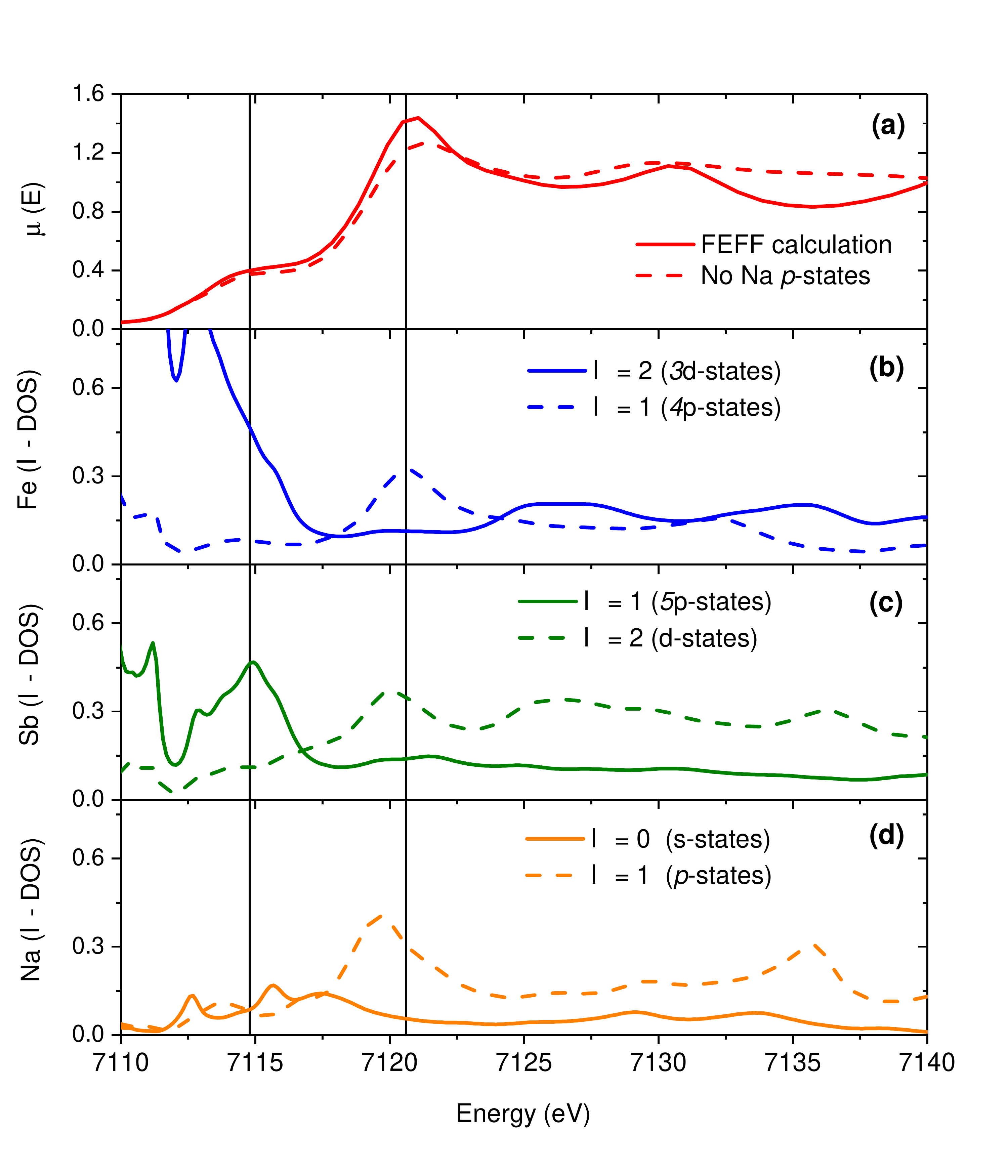}
\par\end{centering}
\caption{(Color online) $(a)$ FEFF calculations of the XANES spectrum and
$(b)-(d)$ site projected $l$-DOS of Na, Fe and Sb for NaFe$_{4}$Sb$_{12}$.
The most relevant contributions are displayed. The main features of
the XANES spectrum correlates well with the Fe $4p$ (general), Sb
$5p$ ($A$ feature), Sb ($d$ $B$ feature) and Na $p$ ($B$ feature)
site projected density of states. The thick line is a guide to the
eyes marking contribution from Sb $5p$ and Fe $4p$ states. \label{fig:calculations_LDOS}}
\end{figure}

The $B$ feature is absent in the Fe foil spectrum and reflects the
Fe coordination structure. Indeed, the measured spectrum is reminiscent
of the Fe $K$-edge spectra in systems containing Fe$^{2+}$ cations
in octahedral symmetry \cite{wilke_oxidation_2001,groot_1s_2009}
and suggests that the Fe atoms in the investigated skutterudites are
close to a $2+$ formal valence state. This result supports the qualitative
discussion on the electronic properties and bond scheme of the $R$Fe$_{4}$Sb$_{12}$
\cite{jeitschko_lafe4p12_1977,noauthor_filled_2010} skutterudites.
A more detailed understanding of the distinct contributions to the
spectrum is attained by comparing the FEFF calculated spectrum with
the site projected local density of states (LDOS) for Fe, Sb and Na.
The results are presented in Figs. \ref{fig:calculations_LDOS}$(a)$-$(d)$. 

The spectrum in Fig. \ref{fig:calculations_LDOS}$(a)$ correlates
well with the Fe $4p$ contribution to the LDOS. Thus, the Fe $K$-edge
can indeed be well described as a Fe $1s\rightarrow4p$ transition
and, furthermore, it indicates that the distortion of the octahedral
coordination allows a certain degree of Fe $3d$-$4p$ mixing that
contribute to the pre-edge region. In addition, it is noteworthy that
the pre-edge contains a large contribution from the Sb $5p$ orbitals
and that the $B$ feature contains large contributions from Sb $d$
and Na $p$ states (or, in general, $R$ $p$ states). 

Turning our attention to the $B$ feature, we performed distinct FEFF
calculations removing the Sb $d$ orbitals and the Na $p$ orbitals.
Removal of the former leads to an overall decrease of the intensity
of the spectrum. The removal of the Na $p$ contribution (as can be
observed) implies a much less intense $B$ feature, without any other
sizable effects. Similar calculations were run for all other fillers
and these results are representative of what is observed for the whole
series. This study allows the conclusion that the FEFF calculation
overestimates the filler $p$ contribution to the $B$ feature. In
all cases, a significant contribution of Sb $5p$ states to the pre-edge
was found, giving extra support for the origin of the pre-edge intensity
of the spectrum as above discussed. Having established the distinct
contributions to the spectrum, we now turn to a comparison between
the experimental results across the series. 

The XAS experimental data in the XANES region of the $R$Fe$_{4}$Sb$_{12}$
series are presented in Fig \ref{fig:XANESallDATA}$(a)$. The lower
inset shows in detail the pre-edge region for all samples. The splitting
of the pre-edge observed for $R=$ Ca is also apparent for the $R=$
Sr or Ba, but is weaker. Again, we resort to the comparison with the
respective FEFF calculations, that will also display a splitting of
about $\Delta E=1.2$ eV, to trust that this weak effect is an intrinsic
property of the system. Thus, the splitting of the pre-edge transition
is indicated to be a characteristic of the non-magnetically ordered
compounds. If we now focus our attention back to Fig. \ref{fig:calculations_LDOS}$(a)$-$(d)$,
a close inspection of Sb $5p$ LDOS suggests that this contribution
is the source of the pre-edge splitting. It is predicted for all $R$Fe$_{4}$Sb$_{12}$,
but is observed only for the alkaline earth filled systems. The result
can be interpreted in terms of the distinct electronic occupation
at $E_{F}$ expected for the $R$Fe$_{4}$Sb$_{12}$ skutterudites,
since the presence of available states in the cases for which $R=$
Ca, Sr or Ba allows the transition to the Fe 3$d$ Sb $5p$ mixed
orbitals close to $E_{F}$. Alternatively, it may relate to distinct
Fe 3$d$ Sb $5p$ mixing when the cases $R=$ Na or K and $R=$ Ca,
Sr or Ba are compared. Therefore, the electronic states derived from
Fe 3$d$ Sb $5p$ mixing are either more occupied or less localized
for the ferromagnetic ordered compounds. 

No shift is observed in the pre-edge position (first inflection),
as it is the case as well of the edge position. The lack of a shift
in the pre-edge does not imply that $E_{F}$ is constant across the
series but rather that the extra electron resulting from alkaline
earth filling will not sit at the Fe site. Indeed, a more involved
scenario is in order, since it is expected that the extra electrons
contributed by the alkaline earth will further stabilize the Sb-Sb
bonds. In general, one should be careful when tracking the destiny
of the extra carriers due to electron or hole doping in itinerant
systems. A similar situation was recently clarified in the case of
the iron arsenides \cite{baledent_electronic_2015}.

In Fig.\ref{fig:XANESallDATA} $(b)$ we present the pre-edge peak
intensities (associated to the $A$ feature) compared with the $\gamma$
coefficient from heat capacity (from Ref. \cite{schnelle_magnetic_2008}).
The upper inset of Fig \ref{fig:XANESallDATA}$(a)$ display a representative
result for the peak fitting procedure, wherein it is shown that only
one Gaussian was adopted to fit the pre-edge transition. 

In view of our interpretation of the pre-edge intensity, the data
is capturing the evolution of the Fe $3d$ Sb $5p$ orbital mixing
and occupation as a function of $R$, both of which are connected
to $\gamma$. It is observed in Fig \ref{fig:XANESallDATA}$(b)$
that, with the exception of the case of CaFe$_{4}$Sb$_{12}$ (see
discussion related to the EXAFS results), the intensities follow qualitatively
the trend of the $\gamma$ coefficient across the series, adding yet
another piece of evidence in support of our interpretation of the
pre-edge peak intensity.

Orbital mixing is strongly influenced, as well, by local structural
parameters, such as Fe - Fe' and Fe - Sb distances and/or the Fe -
Sb - Fe' bond angle (termed $\delta$ in Fig. \ref{fig:structure_properties}).
As shown in Fig. \ref{fig:structure_properties}, the Fe - Fe' distance
is larger for $R=$ Ba while $\delta$ is less, approaching $90$
degrees. Both tendencies would contribute to the suppression of the
$A$ feature intensity of the $R=$ Ba system, as is observed. It
must be appreciated, however, that the relative change of the structural
parameters as a function of $R$ is small and it is not clear if such
small changes could affect the Fe $3d$ Sb $5p$ orbital mixing in
a significant manner. Based on EXAFS results, we turn now to a more
detailed site specific analysis of the structural parameters.

\begin{figure}
\begin{centering}
\includegraphics[scale=0.28]{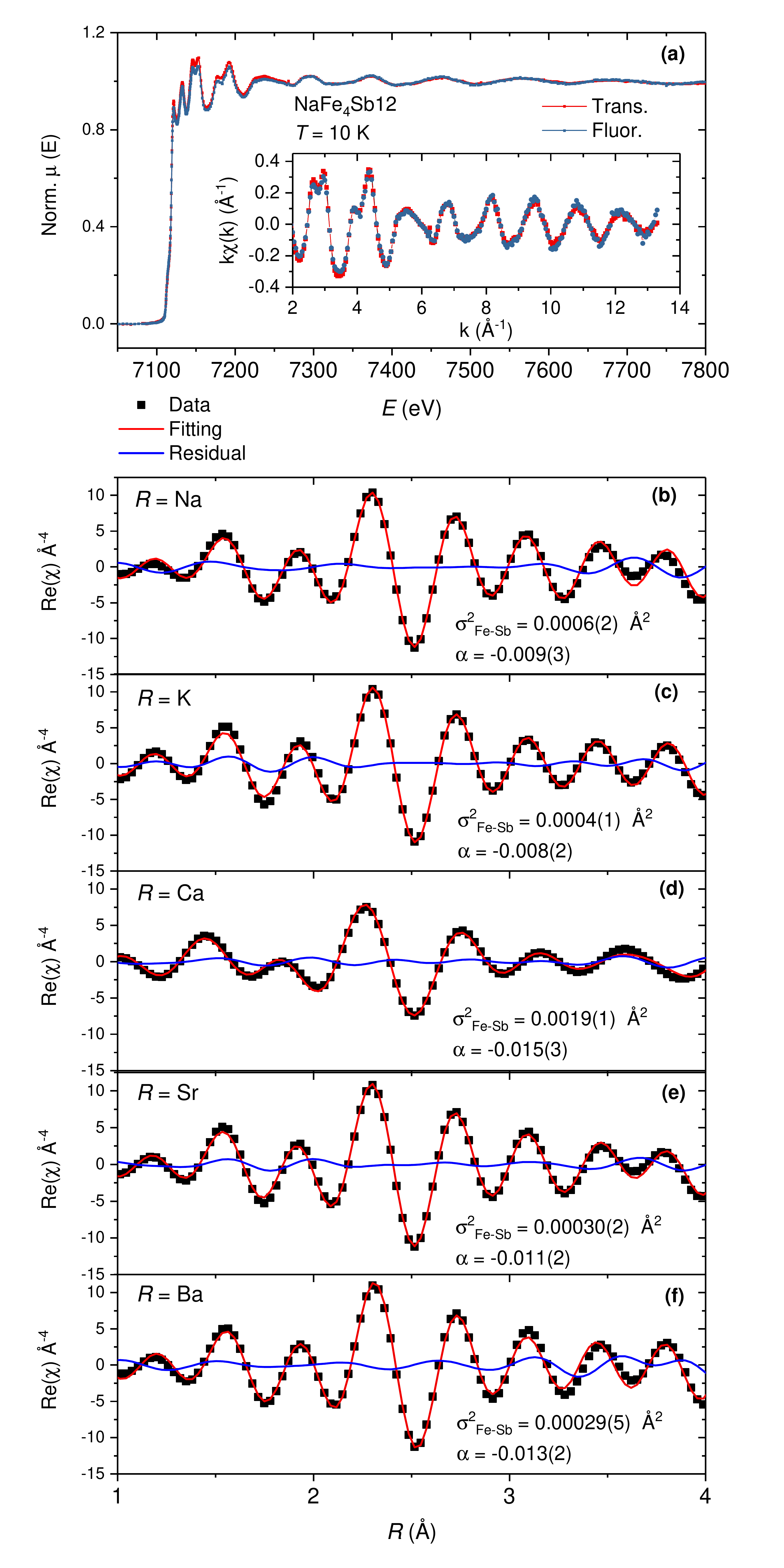}
\par\end{centering}
\caption{(Color online) $(a)$ Representative XAS Fe $K$-edge spectra measured
at $T=10$ K. In the inset, the Fourier transformed spectra in a broad
interval is presented. $(b)-(f)$ Fourier transformed spectra (squares)
up to $4$ $\mathring{\text{A}}$ from the absorber and the theoretical
model (solid red line) used to investigate the Fe coordination structure
along the series ($R=$ Na, K, Ca, Sr, Ba). \label{fig:EXAFSresults}}
\end{figure}

In Fig. \ref{fig:EXAFSresults}$(a)$-$(f)$ we present the results
of our EXAFS investigation at $T=10$ K. The application of EXAFS
to the skutterudites contributed valuable information concerning bond
disorder and the vibrational dynamics in this class of materials \cite{cao_evidence_2004,bridges_complex_2015,noauthor_local_2016}.
In Fig. \ref{fig:EXAFSresults} $(a)$ we present a broad view of
the acquired spectra, comparing the detection from fluorescence and
transmission. In the inset, we show the Fourier transformed spectra
in a broad range. In this paper, as presented in Figs. \ref{fig:EXAFSresults}$(b)$-$(f)$,
our analysis concentrates only in the region up to $4$ $\mathring{\text{A}}$
from the absorber, since we are interested only in evaluating details
of the Fe coordination structure. Single and multiple scattering paths
relating Fe and Sb atoms and a single scattering Fe - $R$ path were
taken into consideration to describe the data. The results for the
obtained $\sigma_{\text{Fe-Sb}}^{2}$ correlated Debye-Waller factors
and for the $\alpha$ parameters adopted to describe the thermal contraction
of the Fe coordination, are presented in the figures. All data could
be well described without taking into account any particular distortion
of the Fe coordination. 

Electronic structure calculations suggest that the Fe derived $t_{2g}$
orbitals provide the main contribution to the system magnetic moments
\cite{leithe-jasper_neutron_2014,xing_magnetism_2015}. Therefore,
Fe - Sb bond disorder is an important parameter in evaluating magnetic
properties since it may affect the degeneracy of the $t_{2g}$ orbitals
and act as a local frustration mechanism of the $t_{2g}$ derived
moments, quenching magnetic order. In principle, the analysis of the
data at $T=10$ K reduces the effects of thermal motion and allows
the Fe- Sb bond disorder to be inferred by comparing the $\sigma_{\text{Fe-Sb}}^{2}$
for distinct fillers. However, the $R$Fe$_{4}$Sb$_{12}$ skutterudites
with light fillers present an unusual vibrational dynamics, as investigated
in detail \cite{koza_breakdown_2008,koza_vibrational_2011}, which
will affect the correlated Debye-Waller factors \cite{bridges_complex_2015},
rendering the comparison across the whole series inadequate. 

A strong indication of this inadequacy is that the correlated Debye-Waller
factors $\sigma_{\text{Fe-\ensuremath{R}}}^{2}$ are larger for the
$R$ = Na, K and Ca than for the $R=$ Sr and Ba compounds, implying
that the Fe - $R$ single scattering paths have little effect on the
$R$ = Na, K and Ca spectra. In this sense, only the $\sigma_{\text{Fe-Sb}}^{2}$
values for $R=$ Na, K and Ca are comparable, while the cases of the
heavy fillers should be taken separately. 

By inspecting the results for $R=$ Na, K and Ca one observes that
$\sigma_{\text{Fe-Sb}}^{2}$ is larger for $R=$ Ca by one order of
magnitude. This bond disorder may act as a local frustration parameter
or, from the standpoint of the Stoner theory, it could simply reduce
the Fe-Fe' exchange and thus the Stoner factor. In any case, this
local parameter is likely more relevant to the suppression of the
ferromagnetic order in CaFe$_{4}$Sb$_{12}$ then the decrease of
the density of states due the change of the filler cation,which is
small when the cases KFe$_{4}$Sb$_{12}$ and CaFe$_{4}$Sb$_{12}$
are compared (Fig. \ref{fig:XANESallDATA}). 

Indeed, the large pre-edge intensity (see Fig\ref{fig:XANESallDATA}$(b)$)
observed for CaFe$_{4}$Sb$_{12}$ connects the local disorder to
local electronic properties, since the large local disorder would
favor the on site Fe $3d$-$4p$ mixing, that is otherwise much too
small due to symmetry constraints. This finding connects the unconventional
vibrational dynamics of skutterudites to its electronic structure,
that in some instances is known to affect their many body electronic
structure \cite{venegas_collapse_2016}.

The absence of order in the cases of filling by Sr or Ba cannot be
clarified in these terms. It could be rooted in the specifics of the
Fe 3$d$ Sb $5p$ mixing and/or orbital occupation that, as suggested
by the pre-edge analysis, is distinct for the $R=$ Na or K and $R=$
Ca, Sr or Ba cases. In the context of the Fe 3$d$ Sb $5p$ hybridization,
an intriguing possibility is offered by the interpretation \cite{kimura_iron-based_2006,kimura_infrared_2007}
that the heavy quasiparticles at $E_{F}$ develop out of a pseudogap,
that is connected to the Kondo screening of the more localized Fe
$d$ states. The Kondo effect, being a many body state, competes with
the magnetic order. 

The EXAFS analysis also indicates that the thermal contraction of
the Fe coordination, as expressed by $\alpha$, is larger for the
alkaline earth filled skutterudites. This is a point of relevance,
since it implies a significant impact on the Fe coordination as a
function of $T$. A more detailed EXAFS investigation will be the
subject of future work. However, an interesting point related to our
results deserves a final word: it is known that the alkaline earth
filled systems CaFe$_{4}$Sb$_{12}$, SrFe$_{4}$Sb$_{12}$ and BaFe$_{4}$Sb$_{12}$
are all strongly renormalized paramagnets, presenting robust ferromagnetic
fluctuations. The bond shortening and increase in disorder observed
in CaFe$_{4}$Sb$_{12}$ is reminiscent of what was observed for SmO$_{1-x}$F$_{x}$FeAs
\cite{cheng_charge_2015}, when quantum criticality is approached.
Thus, the investigation of solid mixtures containing Na or K and Ca,
Sr, or Ba is invited and may unveil some underlying quantum critical
point controlling the physics of the $R$Fe$_{4}$Sb$_{12}$ skutterudites. 

\section{Summary and Outlook}

The electronic structure of the $R$Fe$_{4}$Sb$_{12}$ ($R=$Na,
K, Ca, Sr, Ba) skutterudites were investigated by Fe $K$-edge XAS.
With the support of FEFF calculations, we could interpret the specific
contributions of the $R$, Fe and Sb orbital states to the XAS spectra.
The pre-edge was ascribed mainly to Fe $3d$ Sb $5p$ orbital mixing,
providing experimental evidence for strongly hybridized Fe-Sb states
at $E_{F}$. This interpretation was further supported by the detailed
analysis of the pre-edge intensity, which was shown to correlate with
$\gamma$ across the series, with the exception of the $R=$Ca case. 

EXAFS analysis was carried out to investigate the coordination structure
and no particular distortion of the Fe coordination was found. However,
the Fe - Sb bond disorder for CaFe$_{4}$Sb$_{12}$ was found to be
significant larger than for the $R=$ Na or K cases, suggesting that
the absence of a magnetic ordered state in CaFe$_{4}$Sb$_{12}$ is
due to Fe - Sb bond disorder. Moreover, thermal contraction is much
larger for the $R=$ Ca, Sr or Ba cases. 

In the analysis of the pre-edge transition, there is an indication
that the pre-edge structure splits in two components for the paramagnetic
systems, for which $R=$ Ca, Sr or Ba, but not for the ferromagnetic
ordered skutterudites. In the cases of $R=$ Sr or Ba the effect is
particularly weak. However, comparison with FEFF simulations do support
that the effect is likely intrinsic. The splitting suggests that the
electronic states derived from Fe $3d$ Sb $5p$ orbital mixing are
more populated, or less localized, for the ferromagnetic systems. 

In the context of the phenomenology of itinerant magnets hosting magnetic
moments with weak local moment properties, our work reinforces the
idea that the relevance of local properties, such as the electronic
and coordination structures, may be a common thread to a large set
of itinerant magnets.
\begin{acknowledgments}
The authors acknowledge CNPEM-LNLS for the concession of beam time
(proposal No. $20160180$). The XAFS2 beamline staff is acknowledged
for the assistance during the experiments.  BM Jr. acknowledges the
hospitality and financial support from the Department of Chemical
Metal Sciences at the Max Planck Institute for Chemical Physics of
Solids, Dresden, Germany. We thank A. Amon for his assistance with
the sample XRD characterization. 
\end{acknowledgments}

\bibliographystyle{apsrev4-1}
\bibliography{../References_HardXraysRFe4Sb12}

\end{document}